\documentclass[english,12pt]{article}
\usepackage[T1]{fontenc}
\usepackage[utf8]{inputenc}
\usepackage[english]{babel}
\usepackage[pdftex]{graphicx}
\usepackage[width=\textwidth]{caption}
\usepackage[top=24mm, bottom=24mm, left=24mm, right=24mm]{geometry}
\usepackage{array,float, placeins,flafter, longtable,booktabs,pdflscape}
\usepackage{verbatim}
\usepackage{lmodern}
\usepackage{xr}
\usepackage{mathtools}
\usepackage{multirow}
\usepackage{amsfonts,amsmath,amsthm,amssymb,bm,bbm,dsfont}
\usepackage[utf8]{inputenc}
\usepackage{lmodern}
\usepackage{enumitem}
\usepackage{natbib,color}
\usepackage[pdftex]{graphicx}
\usepackage{subfigure,float,booktabs}
\definecolor{winered}{rgb}{0.5,0,0}
\usepackage[pdfstartview=FitH,  bookmarks=true,  bookmarksopen=true,  breaklinks=true,  colorlinks=true,  linkcolor=blue,anchorcolor=my,  citecolor=blue, filecolor=blue,  menucolor=blue,  urlcolor=myRed]{hyperref}

\numberwithin{equation}{section}
\newtheorem{theorem}{Theorem}[section]

\theoremstyle{definition}

\newtheorem{remark}[theorem]{Remark}
{\theoremstyle{plain}
	\newtheorem{assumption}{Assumption}}
\definecolor{my}{rgb}{0.05,0.05,0.5}
\definecolor{myBlue}{rgb}{.1,.1,.5}
\definecolor{myGreen}{rgb}{0,.4,0}
\definecolor{myRed}{rgb}{.25,0.15,.5}

\definecolor{my}{rgb}{0.05,0.05,0.5}
\newcommand{\cond}{\displaystyle \stackrel{d}{\longrightarrow}}

\newcommand{\conp}{\stackrel{p}{\longrightarrow}}

\renewcommand{\mathbf}[1]{\textbf{\textit{#1}}}

\newcommand{\rmnum}[1]{\romannumeral #1}

\newcommand{\Norm}[1]{\mathcal{N}\left(#1\right)}

\newcommand{\Rmnum}[1]{\expandafter\@slowromancap\romannumeral #1@}
\begin{document}
\title{{Mixed LR-$C(\alpha)$-type tests for irregular hypotheses, \\ general criterion functions and  misspecified models}\thanks{First version: November 29, 2024; This version: June 12, 2025.}
	}
	\author{
		Jean-Marie Dufour\footnote{McGill University,  {\texttt{jean-marie.dufour@mcgill.ca}}}
		\and 
		Purevdorj Tuvaandorj\footnote{York University,
		{\texttt{tpujee@yorku.ca}}}
	}
\maketitle
\begin{abstract}
	This paper introduces a likelihood ratio (LR)-type test that possesses the robustness properties of \(C(\alpha)\)-type procedures in an extremum estimation setting.
	
	The test statistic is constructed by applying separate adjustments to the restricted and unrestricted criterion functions, and is shown to be asymptotically pivotal under minimal conditions. It features two main robustness properties. First, unlike standard LR-type statistics, its null asymptotic distribution remains chi-square even under model misspecification, where the information matrix equality fails. Second, it accommodates irregular hypotheses involving constrained parameter spaces, such as boundary parameters, relying solely on root-\(n\)-consistent estimators for nuisance parameters. When the model is correctly specified, no boundary constraints are present, and parameters are estimated by extremum estimators, the proposed test reduces to the standard LR-type statistic.
	
	Simulations with ARCH models, where volatility parameters are constrained to be nonnegative, and parametric survival regressions with potentially monotone increasing hazard functions, demonstrate that our test maintains accurate size and exhibits good power. An empirical application to a two-way error components model shows that the proposed test can provide more informative inference than the conventional \(t\)-test.
	
\end{abstract}
\thispagestyle{empty}
\newpage
%===================
\pagenumbering{arabic} \setcounter{section}{0} \setcounter{page}{1} 

\section{Introduction \label{sec: Introduction}}
\cite{Neyman(1959)} introduced the $C(\alpha)$ statistic as a versatile approach for testing composite hypotheses, especially in cases where exact optimal tests are unavailable or maximum likelihood estimates (MLEs) are difficult to obtain.

This paper introduces a likelihood ratio-type $C(\alpha)$ test (referred to hereafter as LRC$_\alpha$) for nonlinear hypotheses. The test statistic is constructed using log-likelihood functions that are ``orthogonalized'' against the score functions of nuisance parameters, which account for the direction of estimation error. \citet{Neyman(1959)}'s original $C(\alpha)$ statistic is a score-type test, where a score function associated with the parameter of interest is orthogonalized against the nuisance parameter scores. As a result, the relationship between the proposed statistic and Neyman’s $C(\alpha)$ statistic mirrors the relationship between the classical likelihood ratio (LR) test and the Lagrange multiplier (LM) test.\footnote{In fact, \citet{Neyman(1959)} considers a more general score-type function, referred to as the \emph{Cramér function}.}

LR tests are attractive because they are firmly anchored in likelihood theory, invariant to re-parameterization, enjoy Neyman–Pearson optimality, and—under correct specification—yield a simple chi-square null asymptotic distribution that delivers powerful inference without ad-hoc variance estimates.  This pivotality property, however, relies on the information-matrix equality; when the model is misspecified, the equality fails, the LR statistic ceases to pivot and size distortions may arise.  Our misspecification-robust, \(C(\alpha)\)-style adjustment overcomes this by ``orthogonalising'' the restricted and unrestricted criteria before differencing.  The resulting LRC\(_\alpha\) statistic restores the chi-square null law while retaining the advantages of LR statistics, and it does so under the much weaker requirement that nuisance parameters be merely root-\(n\) consistent—whereas conventional LM, LR, and Wald tests require full asymptotic normality of the estimates \citep{Newey-McFadden(1994), Gourieroux-Monfort(1995b)}.\par 

Accordingly, the LRC\(_\alpha\) test operates under weaker assumptions than standard LR procedures, remains valid in a variety of non-regular problems, and thus provides a valuable addition to the econometrician’s toolkit.  Below, we outline several (not mutually exclusive) scenarios in which \(C(\alpha)\)-type tests—and the proposed LRC\(_\alpha\) test in particular—are especially effective.

\begin{enumerate}[leftmargin=*]
\item Nonnormal distribution -- the nuisance parameter estimator may fail to weakly converge to a random vector, or its asymptotic distribution may deviate from Gaussianity. Such nonnormal distributions can arise, for example, when shrinkage methods are employed to estimate nuisance parameters. In these situations, the $C(\alpha)$ statistic remains applicable, even when the classical Wald, score, and LR statistics may not be directly applicable without appropriate modifications.

A prominent example of a non-regular problem where the $C(\alpha)$ test proves effective is inference after model selection or regularization. For further discussion, see \cite{Chernozhukov-Hansen-Spindler(2015b)}, \cite{Belloni-Chernozhukov-Wei(2016)}, \cite{Belloni-Chernozhukov-Chetverikov-Wei(2018)} and \cite{CCDDHNR(2018)}.

	\item Parameter space -- the $C(\alpha)$ test does not rely on specific assumptions about the parameter space. For instance, the parameters can be subject to inequality constraints \citep{Silvapulle-Sen(2011)}, or the model parameters may lie on the boundary of the parameter space. These constraints can also reflect researchers' prior knowledge, such as box constraints imposed during the optimization of likelihood or other objective functions.
	
	Examples of models with boundary-related issues include GARCH-type models and random coefficient models, where the variance parameters are constrained to be nonnegative.
	
	The $C(\alpha)$ test remains asymptotically pivotal under the null hypothesis, even when the model parameters, or some of their components, lie either in the interior or on the boundary of the parameter space, as long as the nuisance parameter is $n^{1/2}$-consistently estimable. To our knowledge, this robustness property in a general constrained estimation setting has not been previously explored.
	
	\item Computational advantage -- the $C(\alpha)$ statistic offers flexibility in incorporating auxiliary estimators of the nuisance parameters obtained through various estimation methods. These include Bayesian estimators for SVAR models \citep[Chapter 5]{kilian2017}, OLS estimators for ARCH models \citep[Chapter 6]{FrancqZakoian2019}, two-step estimators for sample selection and limited dependent variable models \citep{Wooldridge2010}, and minimum distance or moment-matching estimators for macroeconometric models \citep{dejong2012}. 
	
	Whenever convenient, computationally cheaper or numerically stable estimators for the nuisance parameters can be employed in the $C(\alpha)$ statistics. Classical examples where MLEs are challenging to compute but computationally attractive $M$- and moment estimators are readily available include the Cauchy likelihood and mixture distributions \citep[Chapter 5.7]{vanderVaart(1998)}.
	
	\item Data combination -- in the $C(\alpha)$ statistic, the source from which the nuisance parameter estimate is obtained is not need to be specified as long as the nuisance parameter is estimated with a reasonable precision. For example, the parameter estimates can obtained by data combination 
	\citep{imbens1994,ridder2007} or by calibration \citep{cocci2021} obtained from other data sources. 
\end{enumerate}
%===================================
The literature on \(C(\alpha)\) tests is vast, having developed Neyman’s original score-type approach into a rich body of work. Contributions include \citet{LeCam1956}, \citet{BhatNagnur1965}, \citet{BuhlerPuri1966}, \citet{BartooPuri1967}, \citet{Moran1970,Moran1973}, \citet{Chibisov(1973),Chibisov(1980)}, \citet{Chant(1974)}, \citet{Ray(1974)}, \citet{Singh-Zhurbenko(1975)}, \citet{Foutz(1976)}, \citet{Vorobev-Zhurbenko(1979)}, \citet{Bernshtein(1976),Bernshtein(1978b),Bernshtein(1980),Bernshtein(1980b),Bernshtein(1981)}, \citet{LeCam-Traxler(1978)}, \citet{Neyman(1979)}, \citet{Tarone(1979),Tarone(1985)}, \citet{Tarone-Gart(1980)}, \citet{Wang(1981),Wang(1982)}, \citet{Basawa(1985)}, \citet{Ronchetti(1987)}, \citet{Smith(1987c),Smith(1987)}, \citet{Berger-Wallenstein(1989)}, \citet{Hall-Mathiason(1990)}, \citet{Paul-Barnwal(1990)}, \citet{Wooldridge(1990)}, \citet{Dagenais-Dufour(1991)}, \citet{Davidson-MacKinnon(1991),Davidson-MacKinnon(1993)}, \citet{KocherlakotaS-KocherlakotaK(1991)}, \citet{Dufour-Dagenais(1992)}, \citet{Bera-Yoon(1993)}, \citet{Jaggia-Trivedi(1994)}, \citet{Rao(1996)}, \citet{Bera-Bilias(2001)}, \citet{Pal(2003)}, \citet{Dufour-Valery(2009)}, \citet{Chaudhuri-Zivot(2011)}, \citet{Bontemps-Meddahi(2012)}, \citet{Gu2016}, \citet{DufourTrognonTuvaandorj2017}, \citet{Gu2018}, and \citet{DufourTakano2024}.  Despite this extensive development, the literature has not, to our knowledge, introduced a likelihood-ratio–type counterpart to the \(C(\alpha)\) statistic.

Our main application of the LRC$_\alpha$ test in this paper is inference in the presence of boundary parameters, where properties 1–3 mentioned above are directly relevant.

\citet{Moran1973} and \citet{Chant(1974)} analyze the behavior of $C(\alpha)$ tests in parametric models when a subvector of the tested parameters lies on the boundary of a closed parameter space, as in cases involving null variance parameters or homogeneity restrictions. In contrast, we allow \emph{any} subvector of the model parameters to lie either on or near the boundary of the parameter space.

\citet{Andrews(1999), Andrews(2000), Andrews(2001)} study estimation, bootstrap methods, and tests for models with boundary parameters. However, these papers do not consider $C(\alpha)$-type tests. In a similar context, \citet{Ketz2018} proposes a conditional likelihood ratio-type test. In contrast, the $C(\alpha)$ tests we consider are unconditional, use only chi-square critical values, and do not require a one-step update of the initial estimator, thereby avoiding issues such as negative variance parameter estimates.

\citet{Cavaliere2022} propose a robust bootstrap inference procedure based on an initial shrinkage estimator. In comparison, our asymptotic tests are computationally cheaper and do not involve any tuning parameters.

%============================
In the Monte Carlo simulations, we apply our statistics to test the coefficients of ARCH models and the coefficients of Weibull regression, subject to monotonicity conditions on the underlying hazard function. We find that the $C(\alpha)$ tests have an edge over \cite{Ketz2018}'s test, both in terms of level and power, in the simulations conducted.

In the empirical application, we fit a two-way error components model to rice production data and infer the panel regression coefficients, as well as the variances of the individual and time effects and the idiosyncratic error term. Although this is a benchmark model, formal methods for obtaining confidence intervals that are robust to the degeneracy of variance parameters do not seem to be readily available.
We find that the LRC$_\alpha$ and $t$-ratio-based confidence intervals for the regression coefficients, and the variances of the idiosyncratic error term and individual effects, are comparable and significant. However, while the 95\% LRC$_\alpha$ confidence interval for the variance of the time effect is significant, the 95\% $t$-confidence interval includes $0$ and becomes only borderline significant at the 90\% level.

This paper is organized as follows.  
Section~\ref{sec: framework} sets out the framework, defines the LRC\(_\alpha\) statistic, and discusses its key properties.  
Section~\ref{sec: sub LRC} takes up the important case of subvector testing and provides a heuristic argument for its asymptotic validity.  
Section~\ref{sec: Asy.valid} states the assumptions and presents the main asymptotic result for the LRC\(_\alpha\) statistic.  
Simulation results are reported in Section~\ref{Simulations}.  
Section~\ref{sec: EA} provides an empirical application to an error-components model.  
Section~\ref{Conclusion} concludes.

\paragraph*{Notation.} Let \( B_\varepsilon(\theta) \) denote the open ball centered at \( \theta \) with radius \( \varepsilon > 0 \). The symbol \( \| \cdot \| \) denotes the Frobenius norm for matrices. 
The expression \( q_{1-\alpha}(\chi^2_q) \) denotes the \( (1 - \alpha) \)-quantile of the chi-square distribution with \( q \) degrees of freedom. Let ``CMT'' abbreviate the Continuous Mapping Theorem.
%=========
\section{Likelihood ratio $C(\alpha)$-type test\label{Framework}}\label{sec: framework}
%=============================================================================================
Consider a criterion function \( L_n(\theta, X^{(n)}) \) that depends on a \( d \)-dimensional parameter vector \( \theta \in \Theta \subseteq \mathbb{R}^d \), where \( \Theta \) denotes the parameter space, and an observed sample \( X^{(n)} \) of size \( n \). Let \( \theta_0 \) denote the true parameter vector. The score function and negative Hessian are defined as
\begin{equation}
	S_n(\theta) \equiv \frac{\partial L_n(\theta, X^{(n)})}{\partial \theta}, \quad  
	H_n(\theta) \equiv -\frac{\partial^2 L_n(\theta, X^{(n)})}{\partial \theta \partial \theta'},
	\footnote{Differentiability is assumed for convenience; however, the partial derivatives may alternatively be interpreted as left- or right-hand derivatives, as discussed in \citet[Section 3.3]{Andrews(1999)} and \citet[Chapter 8]{Rockafellar-Wets(1998)}. The general forms of \( S_n(\theta) \) and \( H_n(\theta) \) are formalized in Assumption~\ref{CFRegularityConditions} and Theorem~\ref{prop: nulldist LRCa}.}
\end{equation}
and we hereafter write \(L_n(\theta)\) in place of \(L_n(\theta, X^{(n)})\). The probability limit of \(H_n(\theta_0)\) is denoted by \(H(\theta_0)\), which is assumed to be finite and nonsingular.  Let \( \psi : \Theta \to \mathbb{R}^q \) with \(q < d\) be a continuously differentiable transformation satisfying \( \psi(\theta_0) = \psi_0 \) for some fixed vector \( \psi_0 \).  Its Jacobian is
$
\dot{\psi}(\theta)\equiv \partial \psi(\theta)/\partial \theta'$ (see Assumption~\ref{as: psi}).
We develop an asymptotic \(C(\alpha)\)-type likelihood-ratio test for the null hypothesis
\begin{equation}\label{H0 psi}
	H_0(\psi_0) : \; \psi(\theta) = \psi_0 .
\end{equation}
Let \( \hat{\theta} \) be an unrestricted estimator satisfying \( n^{1/2}(\hat{\theta} - \theta_0) = O_P(1) \), and let \( \tilde{\theta} \) be a restricted estimator satisfying \( n^{1/2}(\tilde{\theta} - \theta_0) = O_P(1) \) and \( \psi(\tilde{\theta}) = \psi_0 \). Natural choices for \( \hat{\theta} \) and \( \tilde{\theta} \) are the unconstrained and constrained extremum estimators:
\begin{equation}\label{eq: EE}
	\hat{\theta}^u \equiv \arg\sup_{\theta\in\Theta}L_n(\theta), \quad \tilde{\theta}^{r} \equiv \arg\sup_{\theta\in\Theta,\ \psi(\theta)=\psi_0}L_n(\theta).
\end{equation}
The \( C(\alpha) \)-type likelihood ratio statistic is defined as
\begin{equation}\label{def: LRCa}
	\mathrm{LRC}_\alpha(\psi_0) \equiv 2n \bigl(L_n^u(\hat{\theta}) - L_n^r(\tilde{\theta})\bigr),
\end{equation}
where
\begin{align}
	L_n^u(\theta) &\equiv L_n(\theta) + \tfrac{1}{2} S_n(\theta)' H_n(\theta)^{-1} S_n(\theta),\label{eq: Lu}\\ L_n^r(\theta) &= L_n(\theta) + \tfrac{1}{2} S_n(\theta)' W_n(\theta) S_n(\theta),\label{eq: Lr}\\
	W_n(\theta) &\equiv H_n(\theta)^{-1} \;-\; H_n(\theta)^{-1}\dot{\psi}(\theta)' 
	\bigl[\dot{\psi}(\theta)\,H_n(\theta)^{-1}\,I_n(\theta)\,H_n(\theta)^{-1}\,\dot{\psi}(\theta)'\bigr]^{-1}
	\dot{\psi}(\theta)\,H_n(\theta)^{-1},\label{eq: W}
\end{align}
with \(I_n(\theta)\), evaluated at \(\tilde\theta\), a consistent estimator of the asymptotic variance \(I(\theta_0)\) of \(n^{1/2}S_n(\theta_0)\).  For \(\alpha\in(0,1)\), the level-\(\alpha\) test rejects the null hypothesis in \eqref{H0 psi} when
\[
\mathrm{LRC}_\alpha(\psi_0) \;\ge\; q_{1-\alpha}(\chi^2_{q}).
\]

\paragraph*{Comparison with LR statistic.}
The LRC\(_\alpha(\psi_0)\) statistic applies two correction terms in \eqref{eq: Lu} and \eqref{eq: Lr} that eliminate the effect of estimation error in the unrestricted and restricted criterion functions. Assume \(I_n(\tilde{\theta}) = H_n(\tilde{\theta})\) and let \(\hat{\theta}\) and \(\tilde{\theta}\) be the unconstrained and constrained extremum estimators defined in \eqref{eq: EE}.  

Assuming interior solutions, for \(\hat{\theta}\), the first-order condition gives \(S_n(\hat{\theta}) = 0\), hence the associated correction term in \eqref{eq: Lu} vanishes. For \(\tilde{\theta}\), the constrained optimization gives 
$S_n(\tilde{\theta}) - \dot{\psi}(\tilde{\theta})'\tilde{\lambda}=0,$ 
where \(\tilde{\lambda}\) is the Lagrange multiplier.  Premultiplying by  
\(\bigl[\dot{\psi}(\tilde{\theta}) H_n(\tilde{\theta})^{-1} \dot{\psi}(\tilde{\theta})'\bigr]^{-1}\!
\dot{\psi}(\tilde{\theta}) H_n(\tilde{\theta})^{-1}\) 
and substituting back yields the nuisance parameter score
\[
S_n(\tilde{\theta}) -
\dot{\psi}(\tilde{\theta})' 
\bigl[\dot{\psi}(\tilde{\theta}) H_n(\tilde{\theta})^{-1} \dot{\psi}(\tilde{\theta})'\bigr]^{-1}
\dot{\psi}(\tilde{\theta}) H_n(\tilde{\theta})^{-1} S_n(\tilde{\theta}) = 0,
\]
so that \(S_n(\tilde{\theta})'W_n(\tilde{\theta}) S_n(\tilde{\theta}) = 0\) and the correction term in \eqref{eq: Lr} also disappears. With both corrections equal to zero, we have
\[
L_n^u(\hat{\theta}) = L_n(\hat{\theta})
\quad\text{and}\quad
L_n^r(\tilde{\theta}) = L_n(\tilde{\theta}),
\]
so that \(\mathrm{LRC}_\alpha(\psi_0)\) reduces to the standard LR statistic.

\paragraph*{Comparison with score-type $C(\alpha)$ statistic.}
It is instructive to compare \( \mathrm{LRC}_\alpha(\psi_0) \) with the standard \( C(\alpha) \) statistic for the restriction in \eqref{H0 psi}:
\begin{align}
	\mathrm{C}_\alpha(\psi_0) &\equiv n\, S_n(\tilde{\theta})^\prime H_n(\tilde{\theta})^{-1} \dot{\psi}(\tilde{\theta})^\prime \left( \dot{\psi}(\tilde{\theta}) H_n(\tilde{\theta})^{-1} I_n(\tilde{\theta}) H_n(\tilde{\theta})^{-1} \dot{\psi}(\tilde{\theta})^\prime \right)^{-1} \dot{\psi}(\tilde{\theta}) H_n(\tilde{\theta})^{-1} S_n(\tilde{\theta}).
	\label{def: CalphaStat}
\end{align}
The test rejects \( H_0(\psi_0) \) if \( \mathrm{C}_\alpha(\psi_0) \geq q_{1-\alpha}(\chi^2_q) \).

The statistic in \eqref{def: CalphaStat}, adapted for nonlinear restrictions, was introduced and studied by \citet{Smith(1987c)} in regular likelihood models, and extended to general estimating equations by \citet{Dufour-Trognon-Tuvaandorj(2016)}. It is based on the ``orthogonalized'' score function \( \dot{\psi}(\tilde{\theta}) H_n(\tilde{\theta})^{-1} S_n(\tilde{\theta}) \), which generalizes the effective score in the nonlinear setting.

By contrast, \( \mathrm{LRC}_\alpha(\psi_0) \) is based on ``orthogonalizing'' the criterion function itself rather than its score. Specifically, the unrestricted log-likelihood \( L_n(\hat{\theta}) \) is adjusted using the score \( S_n(\hat{\theta}) \), while the restricted version \( L_n(\tilde{\theta}) \) is adjusted using the score 
\[
S_n(\tilde{\theta}) - \dot{\psi}(\tilde{\theta})^\prime \left[ \dot{\psi}(\tilde{\theta}) H_n(\tilde{\theta})^{-1} I_n(\tilde{\theta})H_n(\tilde{\theta})^{-1}\dot{\psi}(\tilde{\theta})^\prime \right]^{-1} \dot{\psi}(\tilde{\theta}) H_n(\tilde{\theta})^{-1} S_n(\tilde{\theta}).
\]
Hence, the \( \mathrm{LRC}_\alpha(\psi_0) \) statistic preserves the spirit of Neyman’s \( C(\alpha) \) approach but represents a \emph{likelihood-ratio-type} statistic, distinguishing it from the classical \emph{score-type} \( C(\alpha) \) statistic. When \( I_n(\theta) = H_n(\theta) \), and the extremum estimators satisfy \(S_n(\hat{\theta})=0\) and \(W_n(\tilde{\theta})S_n(\tilde{\theta})=0\), the \( \mathrm{C}_\alpha(\psi_0) \) statistic reduces to the LM statistic, whereas \( \mathrm{LRC}_\alpha(\psi_0) \) reduces to the LR statistic.

Note also that \( \mathrm{LRC}_\alpha(\psi_0) \) is built from \emph{two} estimators—\( \hat{\theta} \) and \( \tilde{\theta} \)—as is customary for LR statistics.  In the artificial extreme case where \( \hat{\theta} = \tilde{\theta} \), the \( \mathrm{LRC}_\alpha(\psi_0) \) and \( \mathrm{C}_\alpha(\psi_0) \) statistics coincide.
\par
\paragraph*{Robustness properties.} 
The \(\mathrm{LRC}_\alpha(\psi_0)\) test enjoys several robustness advantages over the standard LR test. Because \(\theta_0\) may lie on—or arbitrarily close to—the boundary of the parameter space, the terms \(S_n(\hat\theta)\) and \(W_n(\tilde\theta) S_n(\tilde\theta)\) generally do not vanish even when \(\hat\theta\) and \(\tilde\theta\) are extremum estimators and \(I_n(\tilde\theta)=H_n(\tilde\theta)\).  By contrast, if \(\theta_0\) lies in the interior of \(\Theta\), these terms are zero,  \(\mathrm{LRC}_\alpha(\psi_0)\) is equivalent to the usual LR statistic.  The classical LR statistic, however, has a non-pivotal null distribution when parameters lie on the boundary \citep{Andrews(2001)}.  In contrast, \(\mathrm{LRC}_\alpha(\psi_0)\) retains a chi-square null limit (see Theorem~\ref{prop: nulldist LRCa}) under the minimal requirement that \(\hat\theta\) and \(\tilde\theta\) are \(n^{1/2}\)-consistent, and therefore remains pivotal even when \(\theta_0\) is at, or near, the boundary.

More broadly, \(C(\alpha)\)-type tests are designed to be robust under a wide range of constraints on \(\theta\) (e.g., monotonicity, sign, symmetry, or shape restrictions); see \citet{Silvapulle-Sen(2011)} and \citet{Gourieroux-Monfort(1995b)} for examples.  Under mild regularity conditions on \(L_n(\theta)\) and \(\psi(\theta)\), the extremum estimators \(\hat\theta^u\) and \(\tilde\theta^r\) remain \(n^{1/2}\)-consistent even when they are not asymptotically normal \citep{Silvapulle-Sen(2011), Andrews(1999), Gourieroux-Monfort(1995b)}; nevertheless, the resulting \(C(\alpha)\) statistic still converges to a chi-square distribution.  By contrast, Wald, score, and standard LR tests typically require full asymptotic normality and often exhibit complicated, boundary-sensitive limiting behavior. This robustness of the \(C(\alpha)\) methodology appears to have been largely underappreciated.

Finally, the information matrix equality \(H(\theta_0)=I(\theta_0)\) is required for a conventional LR statistic to have an asymptotic chi-square distribution. This equality holds for correctly specified models estimated by maximum likelihood or pseudo-maximum likelihood, and for minimum-distance or GMM criteria, but it may fail under misspecification.  The correction terms
\[
\frac{n}{2} S_n(\hat\theta)' H_n(\hat\theta)^{-1} S_n(\hat\theta)
\quad\text{and}\quad
\frac{n}{2} S_n(\tilde\theta)' W_n(\tilde\theta) S_n(\tilde\theta)
\]
ensure that \(\mathrm{LRC}_\alpha(\psi_0)\) remains asymptotically chi-squared under the null even when the information matrix equality fails; see Theorem~\ref{prop: nulldist LRCa}.

%=================== 
\section{Subvector hypothesis}\label{sec: sub LRC}
%===================
In many applications, one is primarily interested in a low-dimensional subvector of \(\theta\). In this subsection, we therefore specialize our LRC\(_\alpha\) test to the subvector hypothesis \(\theta_1 = \theta_{01}\), treating the remaining components as nuisance parameters.

Let 
$\theta=(\theta_{1}^{\prime}, \theta_{2}^{\prime})^{\prime}$, 
where
$\theta_{1}\in \Theta_{1}\subseteq \mathbb{R}^{d_{1}}$, 
$\theta_{2}\in \Theta_{2}\subseteq \mathbb{R}^{d_{2}}$ and
$d=d_{1}+d_{2}$, with corresponding true parameter vector
$\theta_{0}=(\theta_{01}^{\prime}, \theta_{02}^{\prime})^{\prime}$. Partition the score vector and the Hessian matrix conformably as
\[
S_n(\theta) = \begin{pmatrix} S_{n,1}(\theta) \\ S_{n,2}(\theta) \end{pmatrix}, \quad
H_n(\theta) = \begin{pmatrix}
	H_{n,11}(\theta) & H_{n,12}(\theta) \\
	H_{n,21}(\theta) & H_{n,22}(\theta)
\end{pmatrix}.
\]
We consider testing a restriction on the subvector \( \theta_1 \), treating \( \theta_2 \) as nuisance parameters:
\begin{equation}\label{eq: H0}
	H_0: \theta_1 = \theta_{01}, \quad \theta_2 \in \Theta_2.
\end{equation}
Let \( \hat{\theta} \) and \( \tilde{\theta} = (\theta_{01}', \tilde{\theta}_2')' \) be the unrestricted and restricted extremum estimators, respectively, such that
\[
n^{1/2}(\hat{\theta} - \theta_0) = O_P(1), \quad n^{1/2}(\tilde{\theta}_2 - \theta_{02}) = O_P(1),
\]
under the null hypothesis \eqref{eq: H0}. We first consider the benchmark case in which the information–matrix equality holds, that is,
$H(\theta_0) = I(\theta_0)$ and $H_n(\theta)=I_n(\theta)$.  
To reiterate, this setting still encompasses likelihood, minimum-distance, and GMM criteria.  
In this case, the \( \mathrm{LRC}_\alpha \) statistic simplifies to
\begin{equation}\label{def: LRCa sub}
	\mathrm{LRC}_\alpha(\theta_{01}) \equiv 2n \left( L_n^u(\hat{\theta}) - L_n^r(\tilde{\theta}) \right),
\end{equation}
where
\[
L_n^u(\theta) \equiv L_n(\theta) + \frac{1}{2} S_n(\theta)' H_n(\theta)^{-1} S_n(\theta), \quad
L_n^r(\theta) \equiv L_n(\theta) + \frac{1}{2} S_{n,2}(\theta)' H_{n,22}(\theta)^{-1} S_{n,2}(\theta).
\]
The test rejects the null hypothesis in \eqref{eq: H0} when
$
\mathrm{LRC}_\alpha(\theta_{01}) \geq q_{1-\alpha}(\chi^2_{d_1}).$ A heuristic justification for the asymptotic pivotality of the test proceeds as follows. By a mean-value expansion,
\begin{equation}\label{eq: htheta exp}
	n^{1/2}(\hat{\theta} - \theta_0) = -H_n(\theta_0)^{-1} \left[ n^{1/2} S_n(\hat{\theta}) - n^{1/2} S_n(\theta_0) \right] + o_P(1).
\end{equation}
Applying a second-order Taylor expansion of the log-likelihood around \( \theta_0 \),
\begin{equation}\label{eq: UL exp}
	n L_n(\hat{\theta}) = n L_n(\theta_0) + n S_n(\theta_0)'(\hat{\theta} - \theta_0)
	- \frac{n}{2} (\hat{\theta} - \theta_0)' H_n(\theta_0) (\hat{\theta} - \theta_0) + o_P(1).
\end{equation}
Substituting \eqref{eq: htheta exp} into \eqref{eq: UL exp} and rearranging yields
\begin{equation}\label{eq: UL exp2}
	n L_n(\hat{\theta}) = n L_n(\theta_0)
	+ \frac{n}{2} S_n(\theta_0)' H_n(\theta_0)^{-1} S_n(\theta_0)
	- \frac{n}{2} S_n(\hat{\theta})' H_n(\hat{\theta})^{-1} S_n(\hat{\theta}) + o_P(1).
\end{equation}
Similarly, the restricted adjusted criterion satisfies
\[
n L_n^r(\tilde{\theta}) = n L_n(\theta_0)
+ \frac{n}{2} S_{n,2}(\theta_0)' H_{n,22}(\theta_0)^{-1} S_{n,2}(\theta_0)
+ o_P(1).
\]
Combining these expansions, we obtain
\begin{align*}
	\mathrm{LRC}_\alpha(\theta_{01}) &= n S_n(\theta_0)' H_n(\theta_0)^{-1} S_n(\theta_0)
	- n S_{n,2}(\theta_0)' H_{n,22}(\theta_0)^{-1} S_{n,2}(\theta_0) + o_P(1) \notag \\
	&= n \left[ S_{n,1}(\theta_0) - H_{n,12}(\theta_0) H_{n,22}(\theta_0)^{-1} S_{n,2}(\theta_0) \right]' \notag \\
	&\quad \left[ H_{n,11}(\theta_0) - H_{n,12}(\theta_0) H_{n,22}(\theta_0)^{-1} H_{n,21}(\theta_0) \right]^{-1} \notag \\
	&\quad \left[ S_{n,1}(\theta_0) - H_{n,12}(\theta_0) H_{n,22}(\theta_0)^{-1} S_{n,2}(\theta_0) \right]
	+ o_P(1) \notag \\
	&\overset{d}{\longrightarrow} \chi^2_{d_1}.
\end{align*}
Next, to facilitate comparison with the standard LR statistic, assume that  
\(\hat{\theta}\) and \(\tilde{\theta}\) are extremum estimators satisfying 
\(S_n(\hat{\theta}) = 0\) and \(S_{n,2}(\tilde{\theta}) = 0\), without imposing  
the information matrix equality.  In this case,
\begin{align*}
	&\mathrm{LRC}_\alpha(\theta_{01})
	= 2n \left(L_n(\hat{\theta}) - L_n(\tilde{\theta})\right) \notag\\
	&+n\,S_{n,1}(\tilde{\theta})'  
	\left(
	(H_{n}(\tilde{\theta})^{-1})_{11}-
		(H_{n}(\tilde{\theta})^{-1})_{11}\left[(H_n(\tilde{\theta})^{-1} I_n(\tilde{\theta}) H_n(\tilde{\theta})^{-1})_{11}\right]^{-1}	(H_{n}(\tilde{\theta})^{-1})_{11}
	\right)
	S_{n,1}(\tilde{\theta}),
\end{align*}
where $	(H_{n}(\tilde{\theta})^{-1})_{11}$ and $(H_n(\tilde{\theta})^{-1} I_n(\tilde{\theta}) H_n(\tilde{\theta})^{-1})_{11}$ denote the \(d_1\times d_1\) upper-left blocks of $H_{n}(\tilde{\theta})^{-1}$ and $H_n(\tilde{\theta})^{-1} I_n(\tilde{\theta}) H_n(\tilde{\theta})^{-1}$, respectively. The first term in the above statistic is the usual LR statistic, while the second term is an adjustment that restores the \(\chi^2_{d_1}\) null asymptotic distribution of \(\mathrm{LRC}_\alpha(\theta_{01})\).  When \(I_n(\tilde{\theta}) = H_n(\tilde{\theta})\), the adjustment term vanishes and \(\mathrm{LRC}_\alpha(\theta_{01})\) coincides with the standard LR statistic.

%============================

\section{Asymptotic validity}\label{sec: Asy.valid}

This section presents the main result of the paper, which shows that the LRC\(_\alpha(\psi_0)\) statistic has a null limiting \(\chi^2_q\) distribution. To this end, we maintain the following assumptions.

\begin{assumption}[Continuous differentiability of the parameter transformation]\label{as: psi}
	The transformed parameter $\psi(\theta)\in \mathbb{R}^{q}$ is differentiable 
	with respect to $\theta$ in a neighborhood $B_\delta(\theta_0)$ of $\theta_0$ with Jacobian $\dot{\psi}(\theta_0)\equiv \partial \psi(\theta_0)/\partial\theta'$ of full row rank $q$. 
\end{assumption}
%---------------------------------------------------
\begin{assumption}[Convergence rates of the auxiliary estimates]\label{as: rootnconv}
	There exist auxiliary estimates $\tilde{\theta}$ and $\hat{\theta}$ such that 
	\begin{align*}
		n^{1/2}(\tilde{\theta}-\theta_0)
		&=O_{P}(1)\ \text{and}\ \psi(\tilde{\theta})=\psi_0\ \text{under}\ H_0(\psi_0):\psi(\theta_0)=\psi_0,\notag\\
		n^{1/2}(\hat{\theta}-\theta_0)
		&=O_{P}(1).\notag
	\end{align*}
\end{assumption}
% ==================
\begin{assumption}[Criterion function regularity conditions]\label{CFRegularityConditions}~
	\begin{enumerate}[label=(\alph*)]
		\item\label{RC1} The criterion function ${L}_n(\theta)$ and score type function $S_n(\theta)$ admit the expansions:
		\begin{align}
			L_n({\theta})
			&=L_n({\theta}_0)+S_n(\theta_0)'({\theta}-\theta_0)
			-\frac{1}{2}({\theta}-\theta_0)'H_n(\theta_0)({\theta}-\theta_0)+R_{n}(\theta),\label{eq: Lexp}\\
			S_n({\theta})
			&=S_n({\theta}_0)-H_n(\theta_0)({\theta}-\theta_0)+r_{n}(\theta),\label{eq: Sexp}
		\end{align}
		where the remainder terms satisfy for all $\epsilon>0$
		\begin{equation}\label{eq: rems}
			\sup_{\theta\in\Theta: \Vert n^{1/2}(\theta-\theta_0)\Vert<\epsilon}|R_{n}(\theta)|
			=o_P(n^{-1}),\quad 
			\sup_{\theta\in\Theta: \Vert n^{1/2}(\theta-\theta_0)\Vert<\epsilon}\Vert r_{n}(\theta)\Vert=o_P(n^{-1/2})
		\end{equation}
		\item\label{RC2}  $n^{1/2}{S}_n(\theta_0)\cond {S}\sim \Norm{0, {I}(\theta_0)},$ where 
		${I}(\theta_0)$ is a fixed nonsingular matrix;
		\item\label{RC3} $\sup_{\theta\in\Theta: \Vert\theta-\theta_0\Vert \leq \varepsilon_n}\Vert {H}_n(\theta)-{H}(\theta_0)\Vert=o_{P}(1),$ where 
		${H}(\theta_0)$ is a fixed nonsingular matrix, and $\sup_{\theta\in\Theta: \Vert\theta-\theta_0\Vert \leq \varepsilon_n}\Vert{I}_n(\theta)-{I}(\theta_0)\Vert=o_{P}(1)$ for all $\varepsilon_n\to 0$.
	\end{enumerate}
\end{assumption}
Assumption \ref{as: psi} is standard in the literature; see, for example, \citet{Newey-McFadden(1994)} and \citet{Dufour-Trognon-Tuvaandorj(2016)}. Assumption \ref{as: rootnconv} is also conventional in the context of $C(\alpha)$-type tests.
Assumption \ref{CFRegularityConditions} requires that the criterion function ${L}_n(\theta)$ admits a quadratic expansion, with an asymptotically normal sample score-type function and locally uniformly convergent Hessian and information matrices. This assumption is not restrictive; sufficient conditions are provided in \citet[Section 3.3 and Lemma 1]{Andrews(1999)}.\par 

The main result of this paper is as follows.
%---------------------------------------------------------
\begin{theorem}[The null asymptotic distribution of the LRC$_\alpha(\psi_0)$ statistic]\label{prop: nulldist LRCa}
	Let Assumptions \ref{as: psi}, \ref{as: rootnconv} and \ref{CFRegularityConditions} hold. Then, under \eqref{H0 psi}, as $n\to\infty$
	\begin{align}
		\mathrm{LRC}_{\alpha}(\psi_0)\cond \chi^2_{q}.\label{eq: null dist LRCa}
	\end{align}
\end{theorem} 

\begin{remark}\label{remark: R}~
	\normalfont
	\begin{enumerate}[label={(\arabic{enumi})}, leftmargin=*]
		
		\item\label{R1} To compute the LRC$_\alpha$ statistic, it is necessary to evaluate the criterion (or the  log-likelihood) function \( L_n(\theta) \), the score function \( S_n(\theta) \), the covariance estimator $I_n(\theta)$ and the Hessian \( H_n(\theta) \) at the unrestricted and restricted estimators, \( \hat{\theta} \) and \( \tilde{\theta} \). This is not particularly demanding in practice, as many statistical software packages that implement maximum likelihood or related estimation methods also provide the gradient, Hessian, and the value of the objective function at the final estimates. Furthermore, since the statistic is of \( C(\alpha) \)-type, exact computation of the unconstrained and constrained estimators is not required. It suffices to use any \( n^{1/2} \)-consistent estimators that satisfy the constraints and to evaluate the relevant quantities at these points.
	\item The \(\mathrm{C}_\alpha(\psi_0)\) statistic in \eqref{def: CalphaStat} likewise converges to a chi-square distribution under $H_0(\psi_0)$ and conditions similar to Assumptions \ref{as: psi}, \ref{as: rootnconv} and \ref{CFRegularityConditions}  \citep{DufourTrognonTuvaandorj2017}. 
	In finite samples, however, the \(\mathrm{C}_\alpha(\psi_0)\) and LRC\(_\alpha(\psi_0)\) tests can exhibit markedly different size and power, the latter often performing better; see Section \ref{Simulations}.
	
\item While Theorem \ref{prop: nulldist LRCa} provides an asymptotic justification, one can also implement an exact Monte Carlo version of the test in parametric models \citep{Dufour(2006)} or apply bootstrap methods to the LRC\(_\alpha\) test. Since the statistic requires only \(n^{1/2}\)-consistent estimators, it can be bootstrapped in various ways. The simplest is to resample the data and ``recycle'' the original estimators \(\hat\theta\) and \(\tilde\theta\) when computing the bootstrap statistic.

\item What if \(H_n(\tilde\theta)\) is used in place of \(I_n(\tilde\theta)\)?  If the information matrix equality \(H(\theta_0)=I(\theta_0)\) holds, the LRC\(_\alpha\) statistic retains its standard \(\chi^2_q\) limit.  Otherwise, it converges to a weighted chi-square distribution that is consistently estimable \citep{Hansen2021IE}.  Although this adds complexity relative to the simple chi-square limit in \eqref{eq: null dist LRCa}, robustness to boundary problems is preserved.  

Both bootstrap and Monte-Carlo versions remain valid whether or not \(I_n(\tilde\theta)\) is replaced by \(H_n(\tilde\theta)\).  For a bootstrap LR test that is boundary-robust and avoids the information matrix equality—at the cost of an extra tuning parameter—see \citet{Cavaliere2022}.  A full study of the LRC\(_\alpha\) test with \(I_n(\tilde\theta)=H_n(\tilde\theta)\) and its bootstrap implementation is left for future research.

\end{enumerate}
\end{remark}

\section{Applications to ARCH and survival regression models\label{Simulations}}
\subsection{ARCH model\label{ARCH}}
%===========================================
This section examines the finite sample properties of the bootstrap $C(\alpha)$ tests 
applied to ARCH models. 
The data are generated according to the stationary Gaussian ARCH(4) model
\begin{align*}
x_t&=\sigma_t\epsilon_t,\quad \epsilon_t\sim i.i.d.\, \Norm{0,1},\\
\sigma_t^2
&=\omega+\alpha_1x_{t-1}^2+\dots +\alpha_px_{t-4}^2,\quad t=1,\dots, n.
\end{align*}
The process is initialized at $\{x_{0},\dots, x_{-3}\}=\{0,0,0,0\}$ which are kept fixed throughout the simulations. 
We use the following parameter configurations similar to those considered by \cite{Cavaliere2022}:
\begin{align}
{\text{DGP1}}:&\quad (\omega,\alpha_1,\alpha_2)=(1, 0.1,0),\notag\\
{\text{DGP2}}:&\quad (\omega,\alpha_1,\alpha_2)=(10/9, 0,0),\notag\\
{\text{DGP3}}:&\quad (\omega,\alpha_1,\alpha_2,\alpha_3,\alpha_4)=(1, 0.1,0.1,0.1,0),\notag\\
{\text{DGP4}}:&\quad (\omega,\alpha_1,\alpha_2,\alpha_3,\alpha_4)
=(1, 0.15,0.15,0,0),\notag\\
{\text{DGP5}}:&\quad(\omega,\alpha_1,\alpha_2,\alpha_3,\alpha_4)=(1,0.3,0,0,0,0),\notag\\
{\text{DGP6}}:&\quad(\omega,\alpha_1,\alpha_2,\alpha_3,\alpha_4)=(10/7,0,0,0,0).
\end{align}

We test the hypothesis $H_0: \alpha_2 = 0$ in DGP1-DGP2 and $H_0: \alpha_4 = 0$ in DGP3-DGP6. 
The unconditional variance of $x_t$ is equal to $10/9$ in the first two DGPs and $10/7$ in the remaining designs. The designs DGP1 through DGP6 feature between 1 and 4 boundary parameters, respectively. For instance, in DGP1, the parameter being tested lies on the boundary, while the nuisance parameter $\alpha_1$ is close to the boundary. 
To assess the Type-I error of the tests, we consider the sample sizes $n \in \{250, 500, 1000\}$.

We consider two different estimators for the parameters: the ordinary least squares estimator (OLSE) and the quasi-maximum likelihood estimator (QMLE). The OLSEs are derived from regressing \( x_t^2 \) on a constant and its lags. However, some of these estimates can be negative, violating the stationarity constraint and making the conditional variance, \( \log \sigma_t^2 \), undefined. To address this issue, we use the restricted OLSE approach outlined in \citet[Chapter 6]{FrancqZakoian2019}.

For the QMLE, we fit the ARCH model using the \texttt{rugarch} package in \texttt{R}, employing the \texttt{hybrid} solver, which provides improved convergence. The initial values for the QMLE are both the unrestricted OLSE and the restricted OLSE. The \( C_\alpha \) and \(\text{LRC}_\alpha\) statistics based on these estimators are denoted \( C_{\alpha1} \), \( C_{\alpha2} \), \(\text{LRC}_{\alpha1}\), and \(\text{LRC}_{\alpha2} \). In addition, we implement the test of \citet{Ketz2018}, referred to as \(\text{CLRK}_1\) and \(\text{CLRK}_2\), which are based on a one-step iteration of the QMLE and the restricted OLSE.

The results are presented in Table \ref{tab: ARCHLevel}. The CLRK test consistently underrejects, irrespective of whether the initial estimates for the one-step iteration are based on the QMLE or the restricted OLSE. In contrast, the null rejection rates for the \( C_\alpha \) and \(\text{LRC}_\alpha\) tests are generally accurate, with only minor size distortions. Notably, the rejection rates for these tests improve as the sample size increases. It is also worth highlighting that the \( C_\alpha \) tests based on the restricted OLSE perform particularly well, demonstrating accurate null rejection rates in most, if not all, cases.

\begin{table}[htbp]
\small 
\begin{center}
\caption{Test Type-I Error}
\label{tab: ARCHLevel}
\begin{tabular}{rlllllllllllll}
	\toprule
\multicolumn{2}{c}{ARCH(2)}	 & \multicolumn{2}{c}{C$_\alpha$} &  \multicolumn{2}{c}{C$_{\alpha2}$} &  \multicolumn{2}{c}{LRC$_\alpha$}&  \multicolumn{2}{c}{LRC$_{\alpha2}$} &
\multicolumn{2}{c}{CLRK}&  \multicolumn{2}{c}{CLRK$_2$}\\ 
\cmidrule(lr){1-2}\cmidrule(lr){3-4}\cmidrule(lr){5-6}		\cmidrule(lr){7-8}\cmidrule(lr){9-10}		\cmidrule(lr){11-12}\cmidrule(lr){13-14}
&$n$  & 5\% & 10\% & 5\% & 10\% & 5\% & 10\% & 5\% & 10\% & 5\% & 10\% & 5\% & 10\% \\ 
\cmidrule(lr){2-2}\cmidrule(lr){3-4}\cmidrule(lr){5-6}\cmidrule(lr){7-8}\cmidrule(lr){9-10}		\cmidrule(lr){11-12}\cmidrule(lr){13-14}

DGP1 & 250 &  3.4 &  7.8 &  3.2 &  7.5 &  3.5 &  8.3 &  3.4 &  8.0 &  1.9 &  4.8 &  1.7 &  4.3 \\ 
& 500 &  3.7 &  8.4 &  3.9 &  8.1 &  3.8 &  8.5 &  3.5 &  8.6 &  2.2 &  5.9 &  2.4 &  6.0 \\ 
 & 1000 &  5.0 & 10.8 &  3.6 &  8.2 &  3.8 &  8.3 &  4.6 &  8.9 &  2.1 &  6.2 &  2.3 &  6.6 \\ 
 \midrule 
DGP2 & 250 &  3.8 &  8.2 &  3.8 &  7.6 &  3.4 &  8.2 &  3.8 &  8.7 &  1.4 &  4.2 &  1.5 &  4.2 \\ 
& 500 &  3.8 &  8.3 &  4.6 &  8.8 &  3.8 &  8.2 &  4.8 & 10.7 &  2.5 &  5.7 &  2.4 &  5.7 \\ 
& 1000 &  5.3 &  9.8 &  4.4 &  9.2 &  3.4 &  7.8 &  3.4 &  7.9 &  2.6 &  6.5 &  3.2 &  7.0 \\ 
\end{tabular}
\begin{tabular}{rlllllllllllll}
\toprule
\multicolumn{2}{c}{ARCH(4)}	 & \multicolumn{2}{c}{C$_\alpha$} &  \multicolumn{2}{c}{C$_{\alpha2}$} &  \multicolumn{2}{c}{LRC$_\alpha$}&  \multicolumn{2}{c}{LRC$_{\alpha2}$} &
\multicolumn{2}{c}{CLRK}&  \multicolumn{2}{c}{CLRK$_2$}\\ 
\cmidrule(lr){1-2}\cmidrule(lr){3-4}\cmidrule(lr){5-6}		\cmidrule(lr){7-8}\cmidrule(lr){9-10}		\cmidrule(lr){11-12}\cmidrule(lr){13-14}
&$n$  & 5\% & 10\% & 5\% & 10\% & 5\% & 10\% & 5\% & 10\% & 5\% & 10\% & 5\% & 10\% \\ 
\cmidrule(lr){2-2}\cmidrule(lr){3-4}\cmidrule(lr){5-6}\cmidrule(lr){7-8}\cmidrule(lr){9-10}		\cmidrule(lr){11-12}\cmidrule(lr){13-14}

 DGP3 & 250 &  3.6 &  7.7 &  4.2 &  8.3 &  3.8 &  7.8 &  4.1 &  7.8 &  2.0 &  5.3 &  2.0 &  5.7 \\ 
  & 500 &  4.6 &  8.6 &  4.7 &  9.4 &  4.0 &  8.2 &  4.7 &  8.6 &  2.5 &  6.8 &  2.0 &  4.6 \\ 
  & 1000 &  5.0 &  9.3 &  5.0 & 10.1 &  4.3 &  8.8 &  4.7 &  9.2 &  3.0 &  7.2 &  2.5 &  6.2 \\ 
 \midrule 
 DGP4 & 250 &  4.3 &  8.3 &  3.5 &  7.5 &  3.3 &  7.8 &  2.5 &  6.7 &  1.9 &  6.3 &  2.1 &  4.7 \\ 
  & 500 &  3.3 &  6.9 &  3.7 &  8.6 &  4.0 &  7.6 &  4.1 &  8.7 &  2.2 &  5.9 &  2.0 &  6.2 \\ 
  & 1000 &  4.0 &  9.3 &  3.9 &  8.9 &  4.0 &  8.9 &  4.8 & 10.1 &  2.9 &  6.6 &  2.9 &  6.6 \\ 
 \midrule 
 DGP5 & 250 &  3.1 &  7.0 &  3.8 &  7.7 &  2.9 &  6.6 &  2.9 &  7.3 &  1.8 &  5.9 &  2.1 &  5.1 \\ 
  & 500 &  4.0 &  8.3 &  5.4 &  9.4 &  3.3 &  7.5 &  3.3 &  7.0 &  1.8 &  6.2 &  1.6 &  5.0 \\ 
  & 1000 &  4.8 &  8.6 &  4.5 &  8.9 &  3.2 &  7.3 &  3.9 &  7.8 &  2.8 &  7.1 &  2.2 &  6.2 \\ 

 \midrule 
 DGP6 & 250 &  5.0 &  9.8 &  3.3 &  8.2 &  6.3 & 11.1 &  4.0 &  8.9 &  2.1 &  5.3 &  1.5 &  3.9 \\ 
  & 500 &  4.9 & 10.1 &  3.8 &  8.6 &  7.4 & 12.3 &  4.1 &  8.6 &  2.1 &  6.6 &  1.4 &  4.6 \\ 
  & 1000 &  5.2 &  9.8 &  5.4 & 11.4 &  4.7 & 10.2 &  3.8 &  8.6 &  2.6 &  6.8 &  2.4 &  5.9 \\
 \bottomrule 
	\end{tabular}
	\end{center}
	\footnotesize{\emph{Notes}: C$_\alpha$ and LRC$_\alpha$ are the $C(\alpha)$ tests implemented with QMLE. CLRK denotes the CLR test of \cite{Ketz2018} based on the (unrestricted) QMLE.  
		C$_{\alpha2}$,  LRC$_{\alpha2}$ and CLRK$_2$ are their analogs based on the restricted OLSE. 999 simulations are used to determine the critical values of the CLRK test. The number of replications is 2000.}
\end{table}
%=============================
\subsection{Parametric survival regression}\label{Weibull}
This section presents simulation evidence on the size and power properties of the proposed $\mathrm{LRC}_\alpha$ and related tests. 
We consider a relatively simple design involving a Weibull regresion model, a popular choice for parametric survival analysis \citep[Chapter 17]{CameronTrivedi2009}, featuring a restricted parameter space.  
The data according are generated according to 
\begin{equation}\label{eq: Wbull reg}
	\log t_i=y_i=-\frac{x_i^{\prime}\beta}{\eta}+\frac{u_i}{\eta}\quad i=1,\dots, n,
\end{equation}
where $x_i=(1, x_{1i})^{\prime}$, $x_{1i}\sim i.i.d.\, \mathcal{U}(0,1)$ are kept 
fixed over simulation replications, and the error terms  
$u_i$s are independently drawn from Gumbel distribution with p.d.f. 
$\exp(x-\exp(x))$. $\eta$ is the shape parameter taking values in $\{1,1.01\}$, and the regression coefficients are $\beta=(\beta_0, \beta_1)^{\prime}=(-5, 1)^{\prime}$. 
The hazard function corresponding to \eqref{eq: Wbull reg} is given by  $\lambda(t_i)=\exp(x_i'\beta)\eta t_i^{\eta-1}$, which is monotone increasing in $t_i$ if 
$\eta> 1$. If a researcher imposes this shape restriction a priori, the 
model exhibits a boundary parameter problem.\par 

%===================================================
To elucidate the effect of the restrictions placed on $\eta$ on the test size, we test the joint restriction $H_0:\beta_0=-5, \beta_1=1$ treating 
$\eta$ as nuisance parameter. When estimating the model parameters under the null, two restrictions are considered: (\rmnum{1}) $\eta> 0$,  
(\rmnum{2}) $\eta\geq 1$. We implement the usual parametric score (denoted as LM) and  LR tests, the CLR test of \cite{Ketz2018}, denoted as CLRK, and the $C_\alpha$ and LRC$_\alpha$ tests based on the MLE. The unrestricted MLE is obtained using 
the \texttt{R} package \texttt{survreg} while the restricted estimate of $\nu$ is obtained 
using the \texttt{R} command \texttt{optim} with the \texttt{L-BFGS-B} method.\par 
% Due the simplicity of this optimization problem, the algorithm converged in all cases.\par  
In addition, we consider versions of the $C_\alpha$ and LRC$_\alpha$ tests, denoted as  $C_{\alpha2}$ and LRC$_{\alpha2}$, which use, instead of the restricted MLE, the moment estimator 
\begin{equation}\label{eq: WeibullMomentEst}
	\tilde{\eta}=\frac{-\sum_{i=1}^n(\beta_0+x_{1i}\beta_1)-n\Gamma^{\prime}(1)}{\sum_{i=1}^n\log t_i},
\end{equation}
where $\Gamma^{\prime}(x)$ is the derivative of the gamma function with 
$\Gamma^{\prime}(1)=-0.5772157$. The formula \eqref{eq: WeibullMomentEst} follows from  the first moment of $u_i$. The unrestricted part of the LRC$_{\alpha2}$ test uses the MLE. 

The results are reported in Table \ref{WeibullRegTable}. 
The upper panel corresponds to the case $\eta=1$. When the restriction $\eta \geq 1$ is imposed, the score statistic shows a substantial overrejection, while the LR test is less so but still oversized. 
This observation aligns with the nonpivotality of these statistics when the nuisance parameter is on the edge of the parameter space. However, if one does not assume an increasing hazard function, i.e., the nuisance parameter space is $\eta > 0$, the C$_\alpha$ and LM tests are numerically identical, and the rejection rates of the asymptotic tests are close to the nominal level.

The bottom panel reports the results for the case $\eta = 1.01$. The LR test shows reasonably accurate null rejection rates. The performance of the LM test improves somewhat; however, we still observe some size distortions. This is expected because the nuisance parameter value is now in the interior of the parameter space but still not ``far enough'' from the boundary. On the contrary, the asymptotic $C(\alpha)$-type tests and the CLRK test perform well in line with the theoretical results.
\begin{table}[htbp]
\begin{center}
\caption{Test Type-I Error at $5\%$ level.}
\label{WeibullRegTable}
\begin{tabular}{ccccccccc}
\toprule
$\eta\geq 1$ imposed & $n$	& LM & C$_\alpha$ & C$_{\alpha2}$ & LR & LRC$_\alpha$ & LRC$_{\alpha2}$ & CLRK \\ 
	\midrule 
\multicolumn{9}{c}{$\eta=1$}\\	
\midrule 
Y & 100 & 31.0 &  4.7 &  4.7 &  8.5 &  5.3 &  5.3 &  5.7 \\ 
Y & 250 & 31.4 &  4.3 &  4.3 &  7.4 &  4.3 &  4.7 &  4.0 \\ 
Y & 500 & 32.5 &  4.8 &  4.8 &  8.2 &  5.1 &  5.2 &  5.3 \\ 
N & 100 &  4.9 &  4.9 &  4.7 &  5.3 &  5.3 &  5.3 &  5.7 \\ 
N & 250 &  4.4 &  4.4 &  4.3 &  4.2 &  4.2 &  4.7 &  4.0 \\ 
N & 500 &  5.0 &  5.0 &  4.8 &  5.1 &  5.1 &  5.2 &  5.3 \\ 
	\midrule 
\multicolumn{9}{c}{$\eta=1.01$}\\	
\midrule 
Y & 100 & 17.4 &  4.8 &  4.7 &  6.0 &  5.3 &  5.3 &  5.8 \\ 
Y & 250 & 12.3 &  4.4 &  4.3 &  4.5 &  4.2 &  4.7 &  4.0 \\ 
Y & 500 &  9.0 &  5.0 &  4.8 &  5.3 &  5.1 &  5.2 &  5.1 \\ 
N & 100 &  4.9 &  4.9 &  4.7 &  5.3 &  5.3 &  5.3 &  5.8 \\ 
N & 250 &  4.4 &  4.4 &  4.3 &  4.2 &  4.2 &  4.7 &  4.0 \\ 
N & 500 &  5.0 &  5.0 &  4.8 &  5.1 &  5.1 &  5.2 &  5.1 \\ 
\bottomrule
\end{tabular}
	\end{center}
\footnotesize{\emph{Notes}: LM and LR are the usual score and likelihood ratio tests. C$_\alpha$ and C$_{\alpha2}$ are the $C(\alpha)$ tests implemented with restricted MLE and the moment-based estimator in \eqref{eq: WeibullMomentEst}, respectively. 
LRC$_\alpha$ and LRC$_{\alpha2}$ are the LRC$_\alpha$ tests, both based on the unrestricted MLE but using the restricted MLE and the moment-based estimator in \eqref{eq: WeibullMomentEst}, respectively. CLRK denotes the CLR test of \cite{Ketz2018} implemented with 999 simulations for its critical value. The number of replications is 2000.}

\end{table}
Next, we consider the power of the boundary robust tests. As before, the data are generated according to \eqref{eq: Wbull reg} with the true values 
$(\beta_0,\beta_1,\eta)'=(-5,1,1)$ and $n=100,400$. We test the hypothesis 
$H_0:\beta_0=\beta_0^{*}, \beta_1=1$ where $\beta_0^{*}$ is varied over 
$\{-8,-7.9,\dots, -2.1,-2\}$. We use 499 simulations to obtain the critical value of the CLRK test and the number of replications is 1000.\par
\begin{figure}
\caption{Power of the Boundary Robust Tests}
\label{fig: Power}
		\noindent\makebox[\textwidth]{
			\includegraphics[width=0.6\textwidth]{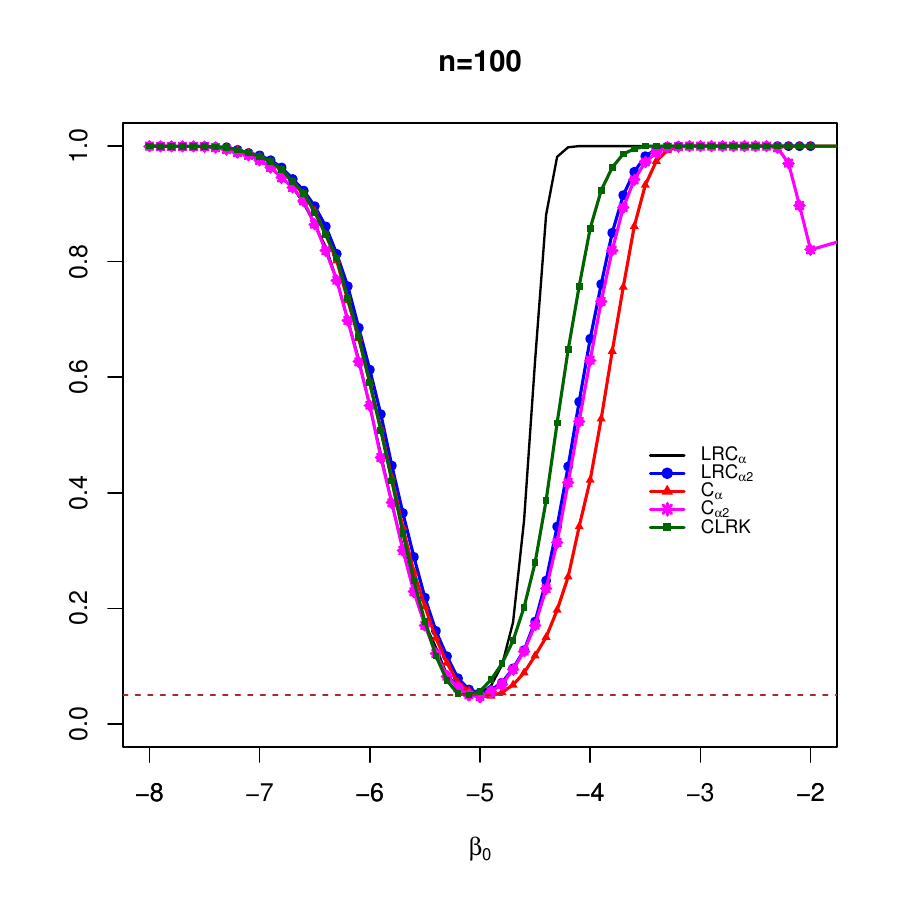}
			\includegraphics[width=0.6\textwidth]{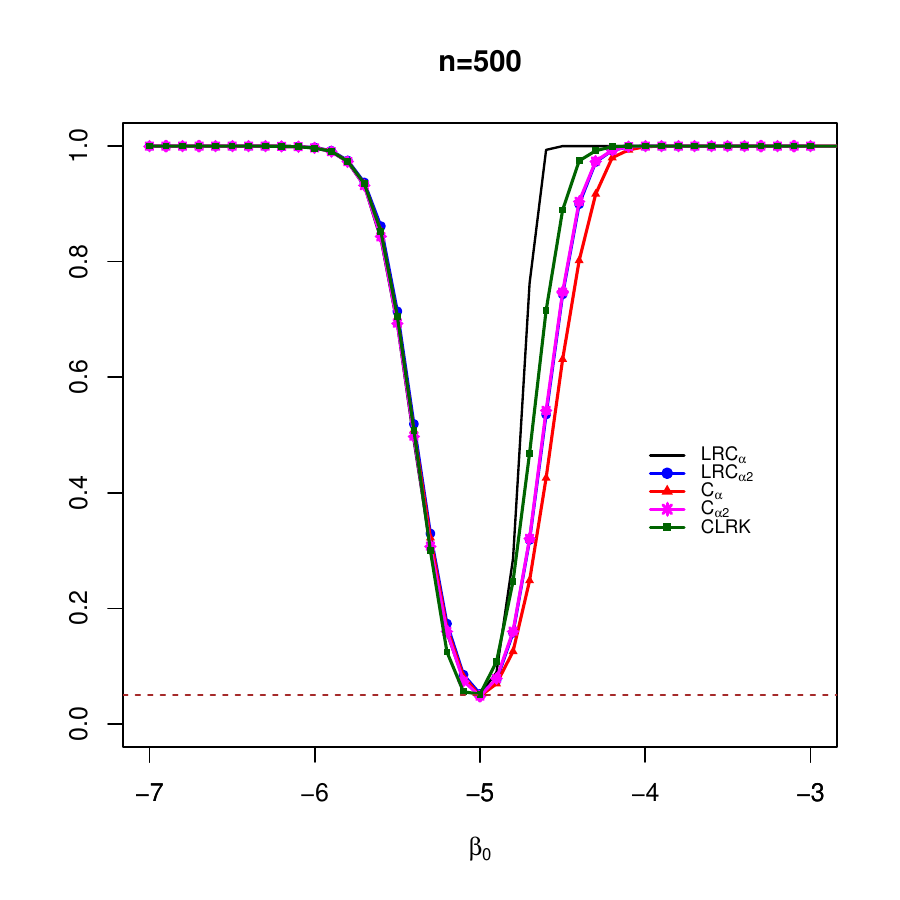}}
		\footnotesize{\emph{Notes}: LM and LR are the usual score and likelihood ratio tests. C$_\alpha$ and C$_{\alpha2}$ are the $C(\alpha)$ tests implemented with restricted MLE and the moment-based estimator in \eqref{eq: WeibullMomentEst}, respectively. 
		LRC$_\alpha$ and LRC$_{\alpha2}$ are the LRC$_\alpha$ tests, both based on the unrestricted MLE but using the restricted MLE and the moment-based estimator in \eqref{eq: WeibullMomentEst}, respectively. CLRK denotes the CLR test of \cite{Ketz2018} implemented with 999 simulations for its critical value. The number of replications is 2000.}
\end{figure}

 
Figure \ref{fig: Power} plots the power curve. The LRC$_\alpha$ and C$_\alpha$ tests that use the restricted MLE are more powerful than the LRC$_{\alpha2}$ and C$_{\alpha2}$
tests based on the moment estimator. Moreover, the C$_{\alpha2}$ shows a non-monotonic power, a common occurrence for (derivative-based) score-type tests. 
The LRC$_\alpha$ test has a higher power than all of the tests, including the CLRK test, for values of $\beta_0$ greater than $-5$, and a nearly equal power for the alternative on its left-hand side.
The null rejection rates of 
the LRC$_\alpha$ and CLRK tests are 
0.053 and 0.0565, respectively when $n=100$, and 0.051
and 0.0515, respectively when $n=500$, so in both cases, the LRC$_\alpha$ test has a more accurate empirical size.

\section{Empirical application to error components model}\label{sec: EA}

This section applies the LRC$_\alpha$ test to construct confidence intervals (CIs) for the variances of the individual, time, and idiosyncratic effects in the two-way error components model, using the \texttt{RiceFarms} dataset from the \texttt{R plm} package. The dataset, analyzed by \cite{horrace1996}, \cite{Druska2004}, \cite{feng2012} and \cite{Croissant2019}, among others, is a panel of $N=171$ rice farms from six villages in West Java, Indonesia, observed over six growing seasons $T=6$ (between 1975 and 1983). These seasons encompass three wet and three dry periods, and the villages vary in terrain and proximity to cities, making it natural to include both farm-specific and time-specific effects in the model specification.

Following \cite{Croissant2019}, we estimate the standard two-way error components regression:

\begin{equation}
	y_{it} = x_{it}'\beta + \eta_i + \lambda_t + v_{it}, \quad i=1, \dots, N; \, t=1, \dots, T,
\end{equation}

where
\begin{itemize}
\item $y_{it}$ is the logarithm of rice output;
\item $x_{it} = (1, \mathrm{log(seed)}_{it}, \mathrm{log(totlabor)}_{it}, \mathrm{log(size)}_{it})'$ is the vector of covariates, where $\mathrm{log(seed)}_{it}$ is the logarithm of the seed used for rice production (in kilograms), $\mathrm{log(totlabor)}_{it}$ is the logarithm of total labor hours (excluding harvest labor), and $\mathrm{log(size)}_{it}$ is the logarithm of the total area cultivated with rice (in hectares).
%\item $x_{it} = (1, x_{1,it}, x_{2,it}, x_{3,it})'$ includes the covariates: $x_{1,it}$ is the log of seed (in kilograms), $x_{2,it}$ is the log of total labor hours (excluding harvest labor), and $x_{3,it}$ is the log of total area cultivated with rice (in hectares);
\item $\{\eta_i\}_{i=1}^N \sim i.i.d. \, (0, \sigma_\eta^2)$ and $\{\lambda_t\}_{t=1}^T \sim i.i.d. \, (0, \sigma_\lambda^2)$ represent unobserved individual-specific and time-specific effects, respectively, while $\{v_{it}\}_{i=1,\dots, N; \, t=1,\dots, T} \sim i.i.d. \, (0, \sigma_v^2)$ are the idiosyncratic error terms.
\end{itemize}
Although various tests exist for the absence of individual-specific and/or time-specific effects \citep[Chapter 4]{Croissant2019}, formal methods for constructing CIs for these variances in the two-way error components model do not appear readily available. Moreover, despite the model's simplicity and widespread use, standard $t$- or Wald-based asymptotic CIs for the variance parameters $\sigma_\eta^2$, $\sigma_\lambda^2$, and $\sigma_v^2$ are not justified due to the nonnegativity constraints imposed on these parameters.

We construct CIs for both the variance parameters and the regression coefficients by inverting the LRC$_\alpha$ test. In addition, we implement $t$-ratio CIs. Both statistics are based on the MLE, which is obtained using the \texttt{maxLik} package in \texttt{R}.

Tables \ref{tab: coefs} and \ref{tab: vars} present the CIs for the regression coefficients and variance parameters, respectively. For the regression coefficients, the LRC$_\alpha$ and $t$ CIs are nearly identical and show significant results, which is unsurprising given that these are linear regression coefficients.

This similarity extends to the CIs for the variances of the idiosyncratic term and the individual effect, both of which show significant results. However, the $t$ CI for the variance of the time effect includes 0 at the 95\% confidence level (with the left endpoint being negative because the CI ignores the nonnegativity constraint), and only narrowly excludes 0 at the 90\% level. In contrast, the robust LRC$_\alpha$ CIs, while slightly wider than the $t$ CIs, produce significant results. Overall, this application illustrates that even in a classical example, discrepancies may arise between the robust and non-robust ($t$) CIs, in which case the robust CI should be preferred.

\begin{table}[ht!]
\begin{center}
\caption{Confidence Intervals for Regression Coefficients}
\label{tab: coefs}
	\begin{tabular}{lcccc}
		\toprule
		&\multicolumn{2}{c}{LRC$_\alpha$}&\multicolumn{2}{c}{$t$}\\
		\cmidrule(lr){2-3}\cmidrule(lr){4-5}
		Variables &  95\% CI  & length &  95\% CI  & length \\ 
		\midrule
		Intercept & [5.39, 6.15] & 0.76 & [5.39, 6.16] & 0.77 \\ 
		log(seed) & [0.15, 0.24] & 0.09 & [0.15, 0.25] & 0.10 \\ 
		log(totlabor) & [0.17, 0.27] & 0.10 & [0.17, 0.28] & 0.11 \\ 
		log(size) & [0.52, 0.62] & 0.10 & [0.51, 0.63] & 0.11 \\ 
		\bottomrule
	\end{tabular}
	\end{center}
	\footnotesize{\emph{Notes}: Both the LRC$_\alpha$ and $t$-ratio CIs are based on the MLE.}
\end{table} 
\begin{table}[ht]
\begin{center}
\caption{Confidence Intervals for Variance Parameters}
\label{tab: vars}
\begin{tabular}{ccccc}
\toprule
&\multicolumn{2}{c}{LRC$_\alpha$}&\multicolumn{2}{c}{$t$}\\
\cmidrule(lr){2-3}\cmidrule(lr){4-5}
Parameters &  95\% CI  & length &  95\% CI  & length \\ 
\midrule 
$\sigma_v^2$ & [0.087, 0.104] & 0.017 & [0.086, 0.104] & 0.018 \\ 
$\sigma_\eta^2$ & [0.015, 0.031] & 0.016 & [0.014, 0.031] & 0.017 \\ 
$\sigma_\lambda^2$& [0.012, 0.128] & 0.116 & [-0.005, 0.068] & 0.073 \\ 
\midrule 	
 &  90\% CI  & length &  90\% CI  & length \\ 
\midrule 
$\sigma_v^2$ & [0.088, 0.102] & 0.014 & [0.087, 0.103] & 0.015 \\ 
$\sigma_\eta^2$  & [0.016, 0.03] & 0.014 & [0.015, 0.029] & 0.014 \\ 
$\sigma_\lambda^2$ & [0.014, 0.098] & 0.084 & [0.001, 0.062] & 0.061 \\ 
\bottomrule
\end{tabular}
\end{center}
\footnotesize{\emph{Notes}: Both the LRC$_\alpha$ and $t$-ratio CIs are based on the MLE.}
\end{table}
%===============================
\section{Conclusion}\label{Conclusion}
The $C(\alpha)$ tests are appealing because $n^{1/2}$-consistent estimates of the nuisance parameters can be obtained through various estimation methods or from different datasets. Moreover, these tests do not require asymptotic normality of the nuisance parameter estimators and impose minimal assumptions on the parameter space.

This paper introduces a novel asymptotic likelihood ratio $C(\alpha)$-type test for general nonlinear restrictions, which appears to be new even in classical testing contexts. While our primary application focuses on testing problems involving boundary parameters—where simulations show the $C(\alpha)$ tests perform well—other potential applications include constrained inference more generally and post-selection inference.

Finally, we aim to extend this work by developing bootstrap versions of the asymptotic $C(\alpha)$ tests to improve their finite-sample performance.

%\newpage
\appendix
%===============================================
%\newpage
\bibliographystyle{chicago}
\bibliography{LRCa}
%\newpage
\section{Proofs}
%============================================
\subsection*{Theorem \ref{prop: nulldist LRCa}}
	By the expansion in \eqref{eq: Sexp},
	\begin{equation}
		S_n(\hat{\theta})
	=S_n({\theta}_0)-H(\theta_0)(\hat{\theta}-\theta_0)+r_{n}(\hat{\theta}).\label{eq: Sexp2}
	\end{equation}
Since $n^{1/2}(\hat{\theta}-\theta_0)=O_P(1)$ by Assumption \ref{as: rootnconv}, for any $\epsilon>0$ and $\delta>0$, there exists $M_0$ large such that $\sup_{n}P[\Vert n^{1/2}(\hat{\theta}-\theta_0)\Vert>M_0]<\epsilon/2$ and $P\left[\sup_{\theta\in\Theta:\Vert n^{1/2}(\theta-\theta_0)\Vert\leq M_0}\Vert n^{1/2} r_n(\theta)\Vert >\delta\right]<\epsilon/2$. Then, using Bonferroni's inequality
\begin{align}
P\left[n^{1/2}\Vert r_n(\hat{\theta})\Vert >\delta\right]
&=P\left[\{n^{1/2}\Vert r_n(\hat{\theta})\Vert >\delta\} \cap\{\Vert n^{1/2}(\hat{\theta}-\theta_0)\Vert\leq M_0\}\right]\notag\\
&\quad+P\left[\{n^{1/2}\Vert r_n(\hat{\theta})\Vert >\delta\} \cap\{\Vert n^{1/2}(\hat{\theta}-\theta_0)\Vert\geq M_0\} \right]\notag\\
&\leq P\left[\sup_{\theta\in\Theta:\Vert n^{1/2}(\theta-\theta_0)\Vert\leq M_0}\Vert n^{1/2} r_n(\theta)\Vert >\delta\right]+\sup_{n}P\left[\Vert n^{1/2}(\hat{\theta}-\theta_0)\Vert\geq M_0 \right]\notag\\
&<\epsilon.\label{eq: Bonferroni}
\end{align}
Since $\epsilon>0$ is arbitrary, we have 
\begin{equation}\label{eq: r=o1}
n^{1/2}\Vert r_n(\hat{\theta})\Vert=o_P(1).
\end{equation}
It follows from \eqref{eq: Sexp2} and \eqref{eq: r=o1} that 
	\begin{align}
		n^{1/2}(\hat{\theta}-\theta_0)&=H(\theta_0)^{-1}n^{1/2}\left[S_n({\theta}_0)-S_n(\hat{\theta})\right]+o_P(1).\label{eq: htheta exp}
	\end{align}
An argument analogous to \eqref{eq: r=o1} yields \( n \vert R_n(\hat{\theta}) \vert = o_P(1) \). Thus, 
by the quadratic expansion in \eqref{eq: Lexp} and Assumption~\ref{as: rootnconv}, we obtain
\begin{equation}
	n L_n(\hat{\theta})
	= n L_n(\theta_0) + n^{1/2} S_n(\theta_0)' n^{1/2}(\hat{\theta} - \theta_0)
	- \frac{1}{2} n^{1/2}(\hat{\theta} - \theta_0)' H(\theta_0) n^{1/2}(\hat{\theta} - \theta_0)
	+ o_P(1). \label{eq: UL exp}
\end{equation}
Upon substituting \eqref{eq: htheta exp} into \eqref{eq: UL exp}, noting that \( n^{1/2} S_n(\theta_0) = O_P(1) \) and \( n^{1/2} S_n(\hat{\theta}) = O_P(1) \), which follows from \eqref{eq: Sexp2}, \eqref{eq: r=o1} and Assumptions~\ref{as: rootnconv} and \ref{CFRegularityConditions}\ref{RC2}, \ref{RC3}, and after some algebra, we arrive at
\begin{align}
	n L_n(\hat{\theta})
	&= n L_n(\theta_0) + \frac{n}{2} S_n(\theta_0)' H(\theta_0)^{-1} S_n(\theta_0)
	- \frac{n}{2} S_n(\hat{\theta})' H(\theta_0)^{-1} S_n(\hat{\theta}) + o_P(1). \label{eq: UL exp2a}
\end{align}
By Assumption~\ref{CFRegularityConditions}\ref{RC3} and a Bonferroni-type argument similar to \eqref{eq: Bonferroni}, we have 
\begin{equation}\label{eq: H con}
H_n(\hat{\theta})\conp  H(\theta_0). 
\end{equation}
Using this and the fact that \( n^{1/2} S_n(\hat{\theta}) = O_P(1) \), we can refine \eqref{eq: UL exp2a} as
\begin{align}
	n L_n(\hat{\theta})
	&= n L_n(\theta_0) + \frac{n}{2} S_n(\theta_0)' H(\theta_0)^{-1} S_n(\theta_0)
	- \frac{n}{2} S_n(\hat{\theta})' H_n(\hat{\theta})^{-1} S_n(\hat{\theta}) + o_P(1). \label{eq: UL exp2 refined}
\end{align}
Therefore,
\begin{equation}
	nL^u_n(\hat{\theta})
	=nL_n({\theta}_0)+\frac{n}{2}S_{n}(\theta_0)'H(\theta_0)^{-1}S_{n}(\theta_0)+o_P(1).\label{eq: UL exp2}
\end{equation}
Next we consider the term $nL^r_n(\tilde{\theta})$. By Taylor expansion and Assumptions \ref{as: psi} and \ref{as: rootnconv}, under the null hypothesis in \eqref{H0 psi}
\begin{equation}\label{eq: psi diff} n^{1/2}(\psi(\tilde{\theta})-\psi(\theta_0))=n^{1/2}\dot{\psi}({\theta}_0)(\tilde{\theta}-{\theta}_0)+o_P(1)=o_P(1).
\end{equation}	
Proceeding similarly to \eqref{eq: Sexp2} and \eqref{eq: H con}, we obtain 
	\begin{align}
	n^{1/2}S_n(\tilde{\theta})
	&=n^{1/2}S_n({\theta}_0)-H(\theta_0)n^{1/2}(\tilde{\theta}-\theta_0)+o_P(1),\label{eq: Sexp3}\\
H_n(\tilde{\theta})&\conp H(\theta_0),\quad I_n(\tilde{\theta})\conp I(\theta_0).\label{eq: HI con}
\end{align}
By the continuity of $\dot{\psi}(\theta)$ at $\theta_0$, $\dot{\psi}(\tilde{\theta})=\dot{\psi}(\theta_0)+o_P(1)$. 
By the CMT, 
$\dot{\psi}(\tilde{\theta})H_n(\tilde{\theta})^{-1}=\dot{\psi}({\theta}_0)H(\theta_0)^{-1}+o_P(1)$.
Combining the latter with \eqref{eq: Sexp3}
	\begin{align}
n^{1/2}\dot{\psi}({\tilde{\theta}})H_n(\tilde{\theta})^{-1}S_n(\tilde{\theta})
&=\dot{\psi}({\theta}_0)H(\theta_0)^{-1}\left[n^{1/2}S_n({\theta}_0)-H(\theta_0)n^{1/2}(\tilde{\theta}-\theta_0)+o_P(1)\right]\notag\\
&=\dot{\psi}({\theta}_0)H(\theta_0)^{-1}n^{1/2}S_n({\theta}_0)
-n^{1/2}\dot{\psi}({\theta}_0)(\tilde{\theta}-{\theta}_0)+o_{P}(1)\notag\\
&=\dot{\psi}({\theta}_0)H(\theta_0)^{-1}n^{1/2}S_n({\theta}_0)+o_{P}(1)\notag\\
&\cond \Norm{0, \dot{\psi}({\theta}_0)H({\theta}_0)^{-1}I({\theta}_0)H({\theta}_0)^{-1}\dot{\psi}({\theta}_0)'},\label{eq: psiHS}
	\end{align}
	where the third equality uses \eqref{eq: psi diff} and the convergence follows from 
	Assumption \ref{CFRegularityConditions}\ref{RC2} and Slutsky's lemma.  Since $\dot{\psi}(\theta_0)$ is of full row rank $q$, by the CMT, 
	\begin{equation}\label{eq: covm con}
	(\dot{\psi}(\tilde{\theta})H_n(\tilde{\theta})^{-1}I_n(\tilde{\theta})H_n(\tilde{\theta})^{-1}\dot{\psi}(\tilde{\theta})')^{-1/2}\conp (\dot{\psi}({\theta}_0)H({\theta}_0)^{-1}I({\theta}_0)H({\theta}_0)^{-1}\dot{\psi}({\theta}_0)')^{-1/2}.
	\end{equation} 
	From \eqref{eq: psiHS}, \eqref{eq: covm con} and Slutsky's lemma,
\begin{equation}\label{eq: psiHS AN}
	(\dot{\psi}(\tilde{\theta})H_n(\tilde{\theta})^{-1}I_n(\tilde{\theta})H_n(\tilde{\theta})^{-1}\dot{\psi}(\tilde{\theta})')^{-1/2}	n^{1/2}\dot{\psi}({\tilde{\theta}})H_n(\tilde{\theta})^{-1}S_n(\tilde{\theta})
\cond \Norm{0,I_q}.
\end{equation}
From \eqref{eq: psiHS AN} and the CMT,
	\begin{equation}
	n\,\Vert	(\dot{\psi}(\tilde{\theta})H_n(\tilde{\theta})^{-1}I_n(\tilde{\theta})H_n(\tilde{\theta})^{-1}\dot{\psi}(\tilde{\theta})')^{-1/2}\dot{\psi}({\tilde{\theta}})H_n(\tilde{\theta})^{-1}S_n(\tilde{\theta})\Vert^2\cond \chi^2_{q}. \label{eq: Calpha(psi) con}
\end{equation}	
We remark here that \eqref{eq: UL exp2 refined} holds when $\hat{\theta}$ is replaced by $\tilde{\theta}$, hence we have 
\begin{align}\label{eq: RL exp2}
nL^r_n(\tilde{\theta})
&=nL_n({\theta}_0)+\frac{n}{2}S_{n}(\theta_0)'H(\theta_0)^{-1}S_{n}(\theta_0)\notag\\
&\quad -\frac{n}{2}
S_n(\tilde{\theta})'H_n(\tilde{\theta})^{-1}\dot{\psi}({\tilde{\theta}})'	(\dot{\psi}(\tilde{\theta})H_n(\tilde{\theta})^{-1}I_n(\tilde{\theta})H_n(\tilde{\theta})^{-1}\dot{\psi}(\tilde{\theta})')^{-1}	\dot{\psi}({\tilde{\theta}})H_n(\tilde{\theta})^{-1}S_n(\tilde{\theta})+o_P(1).
\end{align}
Finally, it follows from \eqref{eq: UL exp2}, \eqref{eq: Calpha(psi) con} and \eqref{eq: RL exp2}
that 
	\begin{align}\label{eq: LRCa limit}
		\mathrm{LRC}_\alpha(\psi_0)
		&=	n\,\Vert	(\dot{\psi}(\tilde{\theta})H_n(\tilde{\theta})^{-1}I_n(\tilde{\theta})H_n(\tilde{\theta})^{-1}\dot{\psi}(\tilde{\theta})')^{-1/2}\dot{\psi}({\tilde{\theta}})H_n(\tilde{\theta})^{-1}S_n(\tilde{\theta})\Vert^2+o_P(1)\notag\\
		&\cond \chi^2_{q}. 
	\end{align}
\end{document}